%% file: paper.tex
\documentclass[12pt]{article}
\usepackage{t1enc}
\usepackage{times}
\usepackage{fullpage}     
\input{makros.tex}

\usepackage{epsfig}
\usepackage{url}
\usepackage{amsmath}
\usepackage{amssymb}
\usepackage{pdflscape}
\usepackage{latexsym}
\usepackage{todonotes}
\usepackage{algorithmic}
\usepackage{algorithm}
\usepackage{multicol}
\usepackage{wrapfig}
\usepackage{numprint}
\npdecimalsign{.} 
\usepackage{theorem}
\usepackage{makeidx}

\newcommand{\ie}{i.e.\ }
\newcommand{\etal}{et~al.~}
\newcommand{\eg}{e.g.\ }

\usepackage{color}
\usepackage{etoolbox}
\patchcmd{\thebibliography}{\list}{\fontsize{0.9em}{0.9\baselineskip}\selectfont\list}{}{} 
\newcommand{\csch}[1]{\color{orange}{\small [CS: #1]}}
\renewcommand{\csch}[1]{}

\newcommand{\mytitle}{Graph Partitioning for Independent Sets}
\begin{document}
\title{\mytitle}
\author{Sebastian Lamm, Peter Sanders and Christian Schulz\\ 
	\textit{Karlsruhe Institute of Technology},
	\textit{Karlsruhe, Germany} \\
	\textit{\normalsize\url{lamm@ira.uka.de}, \{\url{sanders, christian.schulz}\}\url{@kit.edu}}}
\date{}
\maketitle
\begin{abstract}
Computing maximum independent sets in graphs is an important problem in computer science. In this paper, we develop an evolutionary algorithm to tackle the problem.
The core innovations of the algorithm are very natural combine operations based on graph partitioning and local search algorithms.
More precisely, we employ a state-of-the-art graph partitioner to derive operations that enable us to quickly exchange whole blocks of given independent sets. 
To enhance newly computed offsprings we combine our operators with a local search algorithm.
Our experimental evaluation indicates that we are able to outperform state-of-the-art algorithms on a variety of instances. 
\end{abstract}
\thispagestyle{empty}

\section{Introduction}
In a simple and connected graph, an \emph{independent set} is a subset of the nodes such that every pair of nodes that can be formed from the set is not adjacent.
The \emph{maximum independent set} problem is then to find the independent set in the graph with the largest possible cardinality. 
There are lots of applications that benefit from large independent sets such as information retrieval, signal transmission analysis, classification theory, economics, scheduling or computer vision \cite{feo1994greedy}. 
As a more specific example, finding large independent sets is useful in map labeling \cite{gnntbdmlisaac13} where one wants to maximize the number of visible non-overlapping labels on a map. 
Here, a graph model is built such that labels correspond to nodes and there is an edge between two nodes if the associated labels are overlapping. 
It is easy to see that a maximum independent set in the model yields a maximum number of non-overlapping labels.

The maximum independent set problem is closely related to the maximum clique problem and the minimum vertex cover problem.
More precisely, the complement of an independent set $\mathcal{I}$ results in a vertex cover $V\backslash \mathcal{I}$ and an independent set is a clique in the complement graph $\overline{G}$. 
However, note that results from the maximum clique problem are usually only partially transferable to practical algorithms for the maximum independent set problem since building the complement of sparse graphs yields dense graphs. 
It is well known that all of these problems are NP-hard \cite{DBLP:books/fm/GareyJ79}. 
Thus, one relies on heuristic algorithms to find good solutions on large graphs.

Most of the work in literature considers heuristics and local search algorithms for the maximum clique problem (see for example \cite{battiti2001reactive, hansen2004variable,grosso2004combining, katayama2005effective,pullan2006dynamic,grosso2008simple}). 
These algorithms keep a single solution and try to improve it by using node deletions, insertions, swaps as well as the concept of plateau search.
In this context, plateau search only accepts moves that do not change the objective function of the optimization problem.
Heuristics usually employ node swaps to achieve that. 
A node swap refers to the replacement of a node by one of its neighbors; Hence, a node swap cannot directly increase the size of the independent set but can yield a situation where an additional node may get inserted to the solution. 
A very successful approach for the maximum clique problem has been presented by Grosso \etal\cite{grosso2008simple}. In addition to the plateau search approach, different diversification operations are performed and restart rules are added. 
In the independent set context, Andrade \etal\cite{AndradeRW12} extended the notion of swaps to $(j,k)$-swaps. A $(j,k)$-swap removes $j$ nodes from the current solution and inserts $k$ nodes. The authors present a fast linear-time implementation that, given a maximal solution, can find a $(1,2)$-swap or prove that none exists. We implemented the algorithm and use it within our evolutionary algorithm to improve newly computed offsprings.

There are very few papers considering evolutionary algorithms for the maximum independent set problem. 
The general idea behind evolutionary algorithms is to use mechanisms which are highly inspired by biological evolution such as selection, mutation, recombination and survival of the fittest. 
An evolutionary algorithm starts with a population of individuals (in our case independent sets of the graph) and evolves the population into different populations over several rounds. 
In each round, the evolutionary algorithm uses a selection rule based on the fitness of the individuals of the population to select good individuals and combine them to obtain improved offspring \cite{goldbergGA89}. 

B\"ack and Khuri~\cite{back1994evolutionary} and Borisovsky and Zavolovskaya~\cite{borisovsky2003experimental} use fairly similar approaches. They encode solutions as bitstrings such that the value at position $i$ equals one if and only if node $i$ is in the current solution. 
In both cases a classic two-point crossover is used which randomly selects two crossover points $p_1, p_2$. 
Then all bits in between these positions are exchanged between both input individuals.
Note that this likely results in invalid solutions. To guide the search towards valid solutions a penalty approach is used.
A major drawback of the work by B\"ack and Khuri~\cite{back1994evolutionary} is that the authors only test their algorithm on synthetic instances. 
Moreover, in both cases the graphs under consideration are very small. 

The \emph{main contribution} of our paper is a very natural evolutionary framework for the computation of large maximal independent sets. 
The core innovations of the algorithm are combine operations based on graph partitioning and local search algorithms.
More precisely, we employ the state-of-the-art graph partitioner KaHIP~\cite{kaHIPHomePage} to derive operations that enable us to quickly \emph{exchange whole blocks} of given individuals.  
The newly computed offsprings are then improved using a local search algorithm.
In \emph{contrast} to previous evolutionary algorithms, each computed offspring is valid.
Hence, we only allow valid solutions in our population and thus are able to use the cardinality of the independent set as a fitness function.
The rest of paper is organized as follows.\csch{TODO: align with contributions of the paper}
We begin in Section~\ref{s:preliminaries} by introducing basic concepts and  
related work.
We describe the core components of our evolutionary algorithm in Section~\ref{s:evolutionarycomponents}. 
This includes a number of partitioning based combine operators that take two individuals as input 
as well as combine operators that can take \emph{multiple} individuals as input.
A summary of extensive experiments done to tune the algorithm and evaluate its performance is presented in Section~\ref{s:experiments}. 
Experiments indicate that our algorithm computes very good independent sets and outperforms state-of-the-art algorithms on large variety of instances. 
Finally, we conclude with Section~\ref{s:conclusion}.

\section{Prelimiaries}
\label{s:preliminaries}
\subsection{Basic Concepts}
Let  $G=(V=\{0,\ldots, n-1\},E)$ be an undirected graph with $n = |V|$ and $m = |E|$.
The set $N(v)\Is \setGilt{u}{\set{v,u}\in E}$ denotes the neighbors of $v$.
The \emph{complement} of a graph is defined as $\overline{G} = (V,\overline{E})$ with $\overline{E}$ being the complement of $E$. 
An \emph{independent set} is a subset $\mathcal{I} \subseteq V$, such that there are no adjacent nodes in $\mathcal{I}$. 
It is \emph{maximal}, if it is not a subset of any larger independent set.
The \emph{independent set problem} is that of finding the maximum cardinality set among all possible independent sets.
A \emph{vertex cover} is a subset of nodes $C \subseteq V$, such that every edge $e \in E$ is at least incident to one node within the set. The \emph{minimum vertex
    cover problem} asks for the vertex cover with the minimum number of nodes.
It is worth mentioning that the complement of a vertex cover $V \setminus C$ always is an independent set by definition. 
A \emph{clique} is a subset of the nodes $Q \subseteq V$ such that there is an edge between all pairs of nodes from $Q$.

A $k$-way partition of a graph is a division of $V$ into \emph{blocks} of nodes $V_1$,\ldots,$V_k$, \ie $V_1\cup\cdots\cup V_k=V$ and $V_i\cap V_j=\emptyset$
for $i\neq j$.
A \emph{balancing constraint} demands that 
$\forall i\in \{1..k\}\gilt |V_i|\leq L_{\max} := (1+\epsilon)\lceil\frac{|V|}{k}\rceil$
for some imbalance parameter $\epsilon$. 
The objective is to minimize the total \emph{cut} $\sum_{i<j}w(E_{ij})$ where 
$E_{ij}\Is\setGilt{\set{u,v}\in E}{u\in V_i,v\in V_j}$. 
The set of cut edges is also called \emph{edge separator}.
The \emph{$k$-node separator problem} asks to find $k+1$ blocks, $V_1, V_2, \ldots, V_k$ and a separator $S$, that partition $V$, such that there are no edges between the blocks. 
Again, a balancing constraint demands $|V_i| \leq (1+\epsilon)\lceil|V|/k \rceil $. However, there is no balancing constraint on the separator $S$. 
The objective is to minimize the size of the separator $|S|$. 
Note that removing the set $S$ from the graph results in at least $k$ connected components and that the blocks $V_i$ itself do not need to be connected components.
By default, our initial inputs will have unit edge and node weights. 

\subsection{Detailed Related Work}
\label{s:related}
We now discuss algorithmical details of the algorithm by Andrade~\etal\cite{AndradeRW12}.
We call the algorithm ARW as an abbreviation for Andrade, Resende and Werneck. 
While we compare our algorithm against ARW, we also use it within our algorithm to improve newly created offsprings.
Moreover, we shortly present the KaHIP graph partitioning framework since we use it to compute partitions and node separators.

\paragraph*{ARW.}
\label{s:ils}
One \emph{iteration} of the ARW algorithm consists of a perturbation and a local search step.
The ARW \emph{local search} algorithm uses simple $2$-improvements or $(1,2)$-swaps to gradually improve a single current solution. 
A $(j, k)$-swap removes $j$ nodes from the solution and then inserts $k$ new nodes into it. 
A $(1,2)$-swap in particular removes a single node from the solution and adds two other free nodes. 
A node is called \emph{free}, if none of its neighbouring nodes can be found in the current solution. 
The \emph{tightness} of a node $\tau(v)$ is the number of neighbouring solution nodes. 
Hence, free nodes have zero tightness. 
The simple version of the local search algorithm then iterates over all nodes of the graph and looks for a $(1,2)$-swap. 
It is shown, that this procedure can find a valid $(1,2)$-swap in linear time $\mathcal{O}(m)$, if it exists.
This is achieved by using a data structure that allows insertion and removal operations on nodes in time proportional to their degree.
The data structure basically divides the nodes into solution nodes, free nodes and non-free non-solution nodes.
The \emph{perturbation step} used for diversification, forces nodes into the solution and removes neighboring nodes as necessary.
In most cases, one node is forced into the solution per iteration.
With a small probability the number of forced nodes $f$ is set to higher value: $f$ is set to $i+1$ with probability $1/2^i$.
Moreover, the current node to be forced into a solution is picked from a number of random candidates.
Among those candidates the vertex that has been outside the solution for the longest time is picked.
We refer the reader to original paper for more details about the ARW algorithm.
There is also an even faster incremental version of the algorithm that maintains a list of candidates.
We use this version of the algorithm here. 

\paragraph{KaHIP.}
\label{s:kaHIP}
Karlsruhe High Quality Partitioning -- is a family of graph partitioning programs that tackle the balanced graph partitioning problem~\cite{kaHIPHomePage,kabapeE}.  
The algorithms in KaHIP have been able to compute the best results in various benchmarks. 
It implements different sequential and parallel algorithms to compute $k$-way partitions and node separators. 
In this work, we use the sequential multi-level graph partitioner KaFFPa (Karlsruhe Fast Flow Partitioner) to obtain partitions and separators for the graphs. In particular, we use specialized partitioning techniques based on multi-level size-constrained label propagation~\cite{pcomplexnetworksviacluster}.

\section{Evolutionary Components}
\label{s:evolutionarycomponents}
We now discuss the main contributions of the paper. 
We begin by outlining the general structure of our evolutionary algorithm and then explain how we build the initial population.
Finally, we present our new combine operations and the methods we use for mutation.

\subsection{General Structure}
As previous work \cite{back1994evolutionary,borisovsky2003experimental} we use bitstrings as a natural way to represent individuals/solutions in our population.
More precisely, an independent set $\mathcal{I}$ is represented as an array $s = \{0, 1\}^{n}$ where $s[v]=1$ if and only if $v \in \mathcal{I}$.
The general structure of our evolutionary algorithm is very simple.
Our algorithm starts with the creation of a population of individuals (in our case independent sets in the graph) and evolves the population into different populations over several rounds until a stopping criterion is reached. 

In each round, our evolutionary algorithm uses a selection rule that is based on the fitness of the individuals (in our case the size of the independent set) of the population to select good individuals and combine them to obtain improved offspring. 
In contrast to previous work \cite{back1994evolutionary,borisovsky2003experimental}, our combine and mutation operators always create valid independent sets.
Hence, we use the size of the independent set as a fitness function. 
That means that there is no need to use a penalty function to ensure that the final individuals generated by our algorithm are independent sets. 
As we will see later when an offspring is generated it is possible that it is a non-maximal independent set. 
Hence, we apply one iteration of ARW local search without the perturbation step to ensure that it is locally maximal and apply a mutation operation to the offspring.
We use mutation operations since it is of major importance to keep the diversity in the population high \cite{baeckEvoAlgPHD96}, i.e. the individuals should not become too similar, in order to avoid a premature convergence of the algorithm.  

We then use an eviction rule to select a member of the population and replace it with the new offspring. 
In general one has to take both into consideration, the fitness of an individual and the distance between individuals in the population \cite{baeckEvoAlgPHD96}. 
Our algorithm evicts the solution that is \textit{most similar} to the newly computed offspring among those individuals of the population that have a smaller or equal objective than the offspring itself. Once an individual has been accepted into the population we further refine it using additional iterations of the ARW algorithm.
The general structure of our evolutionary algorithm follows the steady-state approach \cite{dejongEvoComp2006} which generates only one offspring per generation. 
We give an outline in Algorithm~\ref{alg:generalsteadystateEA}.

\begin{algorithm}[t]
\begin{algorithmic}
\STATE   \quad create initial population $P$ 
\STATE   \quad \textbf{while} stopping criterion not fulfilled 
\STATE   \quad \quad \textit{select} parents $\mathcal{I}_1, \mathcal{I}_2$ from $P$ 
\STATE   \quad \quad \textit{combine} $\mathcal{I}_1$ with $\mathcal{I}_2$ to create offspring $O$
\STATE   \quad \quad \textit{ARW local search}+\textit{mutation} on offspring $O$ 
\STATE   \quad \quad \textit{evict} individual in population using $O$ 
\STATE   \quad \textbf{return} the fittest individual that occurred
\end{algorithmic}
\caption{Steady State Evolutionary Algorithm with Local Search}
\label{alg:generalsteadystateEA}
\end{algorithm}

\subsection{Initial Solutions}
\label{ss:initialsolutions}
We use three different approaches to create initial solutions. 
Each time we create an individual for the population we pick one of the approaches uniformly at random.
The first and most simplistic way is to start from an empty independent set and add nodes at random until no further nodes can be added. 
To ensure that adding a node results in a valid independent set we have to check if the node is free. 
We do this by simply checking if any of the surrounding nodes is already in the set.
The method adds a decent amount of diversity during the construction phase, which over an extended period of time can lead to good solutions. 

Secondly, we use a greedy approach similar to Andrade~\etal\cite{AndradeRW12}.
Starting from an empty solution, we always add the node with the least residual degree which is the number of free neighbors.
After a node is added to the solution, we remove all its neighbouring nodes from the graph and update the residual degree of their neighbors.
We repeat the procedure until no further node can be added.
The implementation is done using a simple bucket priority queue which groups nodes into buckets based on their residual degree. 
This allows us to pick a random node each time multiple nodes share the same residual degree. 

The last approach that we use to create initial solutions is also a greedy one.
Here, we take a detour and generate an independent set by computing a vertex cover. 
We first create a vertex cover and then compute its complement to get an independent set. 
The algorithms also starts with an empty solution and then always adds the node that will cover the most currently uncovered edges.
We repeat this until all edges are covered and then return the corresponding independent set.
Note that the two greedy algorithms compute different independent sets (\eg consider a path with five nodes). 
While the first approach always maintains an independent set and tries to improve it, the second approach can only return an independent set once the algorithm has terminated.

\subsection{Combine Operations}
We perform different kinds of combine operations which are all based on graph partitioning. 
The main idea of our operators is to use a partition of the graph to exchange whole blocks of solution nodes.
In \emph{general} our combination operators try to generate new independent sets that are not necessarily maximal. We then perform a maximization step that adds as many free nodes as possible. Afterwards, we apply a single iteration of the ARW local search algorithm to ensure that our solution is locally optimal.
Depending on the type of the operator, we use a node separator or an edge separator of the graph that has been computed by the graph partitioning framework KaHIP.
As a side note, small edge or node separators are vital for our combine operations to work well. 
This is due to the fact that large separators in the combine operations yield offsprings that are far from being maximal.
Hence, the maximization step performs lots of fixing and the computed offspring is not of high quality.
This is supported by experiments presented in Section~\ref{s:mainresults}.

The first and the second operator need precisely two input solutions while our third operator is a multi-point combine operator -- it can take multiple input solutions. In the first case, we use a simple tournament selection rule \cite{Miller95geneticalgorithms} to determine the inputs, \ie $\mathcal{I}_1$ is the fittest out of two random individuals $r_1, r_2$ from the population. 
The same is done to select $\mathcal{I}_2$. 
Note that due to the fact that our algorithms are randomized, a combine operation performed twice using the same parents can yield a different offspring. 

\begin{figure}[t]
\begin{center}
\includegraphics[width=5cm]{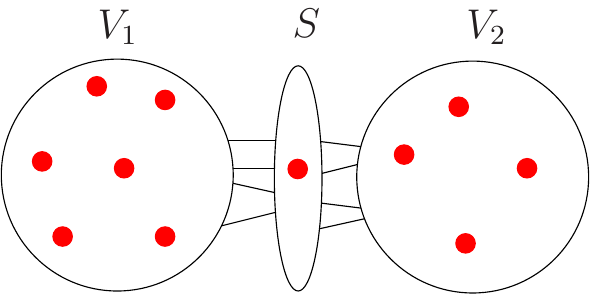} \hspace*{.5cm} 
\includegraphics[width=5cm]{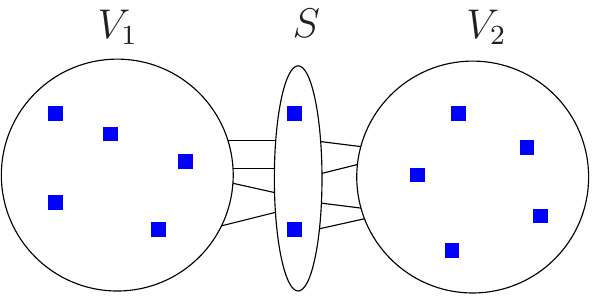}  \\\vspace*{.25cm}
\includegraphics[width=5cm]{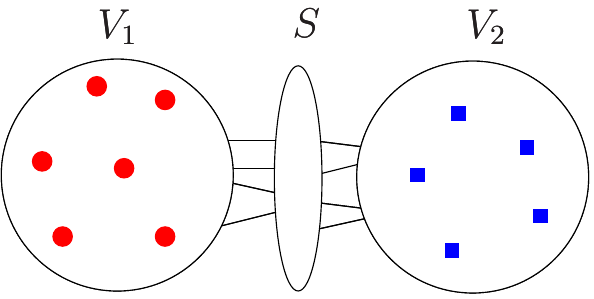} \hspace*{.5cm} 
\includegraphics[width=5cm]{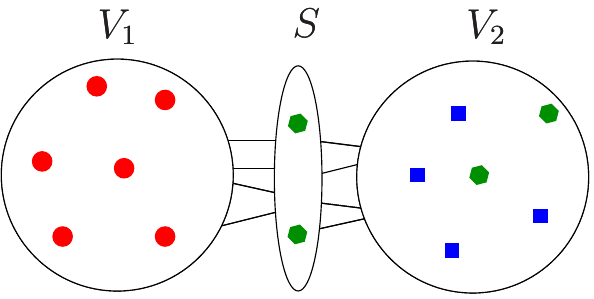}  \\
\end{center}
\caption{An example combine operation using a node separator $V=V_1 \cup V_2 \cup S$. On top two input individuals/independent sets, $\mathcal{I}_1$ and $\mathcal{I}_2$, are shown. Bottom left: a possible offspring that uses the independent set of $\mathcal{I}_1$ in block $V_1$ and the independent set of $\mathcal{I}_2$ in block $V_2$. Bottom right: the improved offspring after ARW local search has been applied to improve the given solution and to add nodes from the separator to the independent set.}
\label{fig:nodesparatorcombine}
\end{figure}

\paragraph{Node Separator Combination.} In its simplest form, the operator starts by computing a node separator $V=V_1 \cup V_2 \cup S$ of the input graph. 
We then use $S$ as a crossover point for our operation.
The operator generates two offsprings.\csch{try whether it is enough to only look at the better one, can we safe time and then get even better solutions in the end}
More precisely, we set $O_1=(V_1\cap \mathcal{I}_1) \cup (V_2\cap\mathcal{I}_2)$ and $O_2=(V_1\cap \mathcal{I}_2) \cup (V_2\cap\mathcal{I}_1)$. 
In other words, we exchange whole parts of independent sets from the blocks $V_1$ and $V_2$ of the node separator. 
Note that the exchange can be implemented in time linear in the number of nodes.
Recall that the definition of a node separator implies that there are no edges running between $V_1$ and $V_2$.
Hence, the computed offsprings are independent sets, but may not be maximal since separator nodes have been ignored and potentially some of them can be added to the solution.
We maximize the offsprings by using the greedy independent set algorithm from Section~\ref{ss:initialsolutions}. 
The operator finishes with one iteration of the ARW algorithm to ensure that we reached a local optimum and to add some diversification.
An example illustrating the combine operation is shown in Figure~\ref{fig:nodesparatorcombine}.

\paragraph{Edge Separator Combination.}
This operator computes offsprings by taking a detour over vertex covers.
It starts by computing a bipartition $V=V_1 \cup V_2$ of the graph.
Let $C_i$ be the vertex cover $V\backslash \mathcal{I}_i$. 
We define temporary vertex cover offsprings similar to before: $D_1=(C_1 \cap V_1) \cup (C_2 \cap V_2)$ and $D_2=(C_1 \cap V_2) \cup (C_2 \cap V_1)$. 
Unfortunately, it is possible that an offspring created this way contains some non-covered edges. These edges can only be a subset of the cut edges of the partition.\csch{todo: give an example, prove claim?}
We want to add as little nodes as possible to our solution to fix this.
Hence, we add a minimum vertex cover of the bipartite graph induced by the non-covered cut edges to our vertex cover offspring. 
The minimum vertex cover in a bipartite graph can be computed using the Hopcroft-Karp algorithm.
Afterwards, we transform the vertex cover back to an independent set, and follow our general approach by applying ARW local search to reach a local optimum. 

\paragraph{Multi-way Combination.}
Our last two operators are multi-point crossover operators that extend the previous two operators.
Both of them divide the graph into a number of blocks $k$. 
Depending on the type of the operator, a node or edge separator is used.
We start with the description of the node separator approach where $V=V_1 \cup \ldots \cup V_k \cup S$.
The operator selects a number of parents. 
We then calculate the score for \emph{every} possible pair of a parent $\mathcal{\mathcal{I}}_i$ and a block $V_j$. 
The score of a pair is  the number of the parents solution nodes inside the given block. 
We then select the parent with the \emph{highest score} for each of the blocks to compute the offspring. 
As before, since we left out the separator nodes we use a maximization step to make the solution maximal and afterwards apply ARW local search to ensure that our solution is a local optimum.

If we use an edge separator for the combination, we start with a $k$-way partition of the nodes $V=V_1 \cup \ldots \cup V_k$.
This approach also computes scores for each pair of parent and block.
This time the score of a pair is defined as the number of the vertex cover nodes of the complement of an independent set inside the given block. 
We then select the parent with the \emph{lowest score} for each of the blocks to compute the offspring. 
As in the simple vertex cover combine operator, it is possible that some cut edges are not covered. 
We use the simple greedy vertex cover algorithm to fix the offspring since the graph induced by the non-covered cut edges is not bipartite anymore.
We then once again complement our vertex cover to get our final offspring. 

\subsection{Mutation Operations}
After we performed a combine operation, we apply a mutation operator to introduce further diversification. 
Previous work~\cite{back1994evolutionary,borisovsky2003experimental} uses bit-flipping for mutation, \ie every bit in the representation of a solution has a certain probability of being flipped. 
We can not use this approach since our population only allows valid solutions. 
Instead we perform forced insertions of new nodes into the solution and remove adjacent solution nodes if necessary as in the perturbation routine of the ARW algorithm. 
Afterwards we perform ARW local search to improve the perturbed solution.

\subsection{Miscellanea}
Instead of computing a new partition for every combine operation, we hold a pool of partitions and separators that is computed in the beginning. A combine operation then picks a random partition or node separator from the pool. If the combine operations have been unsuccessful for too many iterations, we compute a fresh set of partitions. 
In our experiments we used two-hundred unsuccessful combine operations as a threshold.
Additionally, we have to ensure that the partitions created for the combine operations are sufficiently different over multiple runs.  
However, although KaHIP is a randomized algorithm, small cuts in a graph may be similar. 
To avoid similar cuts and increase diversification of the partitions and node separators, we additionally  give KaHIP a random imbalance $\epsilon \in_\text{rnd} [0.05, 0.75]$ to solve the partitioning problem. 
Additionally, we tried one more combine operator based on set intersection. 
This operator computes an offspring by keeping the nodes that are in both inputs which is by definition an independent set. However, our experiments with the operator did not yield good results so that we omit further investigations here.
\section{Experimental Evaluation}
\label{s:experiments}
\paragraph*{Methodology.} 
We have implemented the algorithm described above (EvoMIS) using C++ and compiled all algorithms using gcc 4.63 with full optimization's turned on (-O3 flag). 
We mainly compare our algorithm against the ARW algorithm since it has a relatively clear advantage in Resende \etal\cite{AndradeRW12}.
The algorithm by Grosso \etal\cite{grosso2008simple} has originally been formulated for the maximum clique problem.
Andrade \etal\cite{AndradeRW12} used an implementation of the algorithm for the maximum independent set problem.
Hence, we also compare against the results of the algorithm by Grosso \etal presented in the paper of Andrade \etal\cite{AndradeRW12}.
Additionally, we compare ourselves with our implementation of the evolutionary algorithm presented by B\"ack and Khuri~\cite{back1994evolutionary}.

Unless otherwise mentioned, we perform five repetitions where each algorithm that we run gets ten hours of running time to compute a solution. 
Each run was made on a machine that is equipped with two Quad-core Intel Xeon processors (X5355) which run at a clock speed of 2.667 GHz. 
It has 2x4 MB of level 2 cache each, 64 GB main memory and runs Suse Linux Enterprise 10 SP 1.  
We used the fastsocial configuration of the KaHIP v0.6 graph partitioning package~\cite{kaHIPHomePage} to obtain graph partitions and node separators.
The test results for the ARW algorithm were obtained by using the original algorithm from Andrade \etal\cite{AndradeRW12}.
Within the evolutionary algorithm we used our own implementation of the ARW algorithm.

We mostly present two kinds of data: maximum values, average values, minimum values as well as plots that show the evolution of solution quality.  
We now explain how we compute the convergence plots.
Whenever an algorithm creates a new best independent set $S$ it reports a tuple ($t$, $|S|$), where the time stamp $t$ is the currently elapsed time and $|S|$ refers to the size of the independent set that has been created.
Since we perform multiple repetitions, the final plots correspond to average values over these repetitions.
To compute these we take the time stamps of all repetitions and sort them in ascending order. 
For each time stamp in this series, we report the average value of the best solution size of each repetition at that time.

\paragraph*{Algorithm Configuration.}
After an extensive evaluation of the parameters \cite{baLamm}, we fixed the population size to two hundred fifty, the partition pool size to thirty, the number of ARW iterations to $\numprint{15000}$ as well as the number of blocks used for the multi-way combine operations to sixty-four. In each iteration, one of our three combine operations is picked uniformly at random.
However, our experiments indicate that our algorithm is not too sensitive about the precise choice of the parameters. 
We mark the instances that have also been used for the parameter tuning in \cite{baLamm} in Appendix~\ref{sec:appendixdetailedresults} with a *.
\paragraph*{Instances.}
We use graphs from various sources to test our algorithm. We divide them into five categories: social networks, meshes, road networks, networks from finite element computations as well as networks stemming from matrices. 
Social networks include citation networks, autonomous systems graphs or web graphs taken from the 10th DIMACS Implementation Challenge benchmark set~\cite{benchmarksfornetworksanalysis}. 
 Road networks and meshes are taken from Andrade~\etal\cite{AndradeRW12} and have been kindly provided by Renato Werneck. 
Meshes are dual graphs of triangular meshes. 
Networks stemming from finite element computations have been taken from Chris Walshaw's benchmark archive~\cite{soper2004combined}.
Graphs stemming from matrices have been taken from the Florida Sparse Matrix Collection~\cite{UFsparsematrixcollection}. 
We randomly selected one from each group of all real, symmetric matrices having between 10K and 65K columns.
A graph is derived by inserting a node for each column and creating an edge between two nodes $u, v$ if the corresponding matrix entry is non-zero.
Self-loops are removed from the graphs.

\subsection{Main Results}
\label{s:mainresults}
We now shortly summarize the main results of our experiments. 
First of all, in 50 out of the 67 instances, we either improve or reproduce the maximum result computed by the ARW algorithm.
Our algorithm computes a maximum solution that is strictly larger than the maximum solution computed by the ARW algorithm in 21 cases.
Contrarily, in 17 cases the maximum result of the ARW algorithm is larger then the maximum result of our algorithm.
When looking at average values, we get 23 cases in which our algorithm strictly outperforms the ARW algorithm, and 17 cases for the opposite direction.
Remarkably, when looking at the graphs obtained from the Florida Sparse Matrix collection, the average value of the ARW algorithm only outperforms our algorithm on one instance.  
The mesh family that we use in this paper has also been used in the original ARW paper~\cite{AndradeRW12}. 
We like to stress that most of the maximum results of the ARW algorithm are strictly larger than the maximum values originally reported by Andrade \etal\cite{AndradeRW12} (including the maximum values presented there of the algorithm by Grosso \etal\cite{grosso2008simple}). 
Except for four instances the same holds for our algorithm. 
On these four instances, our algorithm is worse than the original maximum value of the ARW algorithm.
On the mesh family, in 8 out of 14 cases our algorithm computes the best result ever reported in literature.
On road networks and the largest graphs from the mesh family as well as Walshaw family the ARW algorithm outperforms our algorithm. 
We tried to give both algorithms more time, \ie a whole day of computation, but did not see much different results.
Lastly, there is an interesting observation on social networks, that is in 5 out of 9 cases the minimum, average and maximum result produced by both algorithms are precisely the same. 
We suspect that these instances are in a sense easy and that both algorithms compute the optimal result or are very close to the optimum.
We provide detailed per instance results in Appendix~\ref{sec:appendixdetailedresults}.
\begin{figure}[t]
	\centering
	\includegraphics[width=8cm]{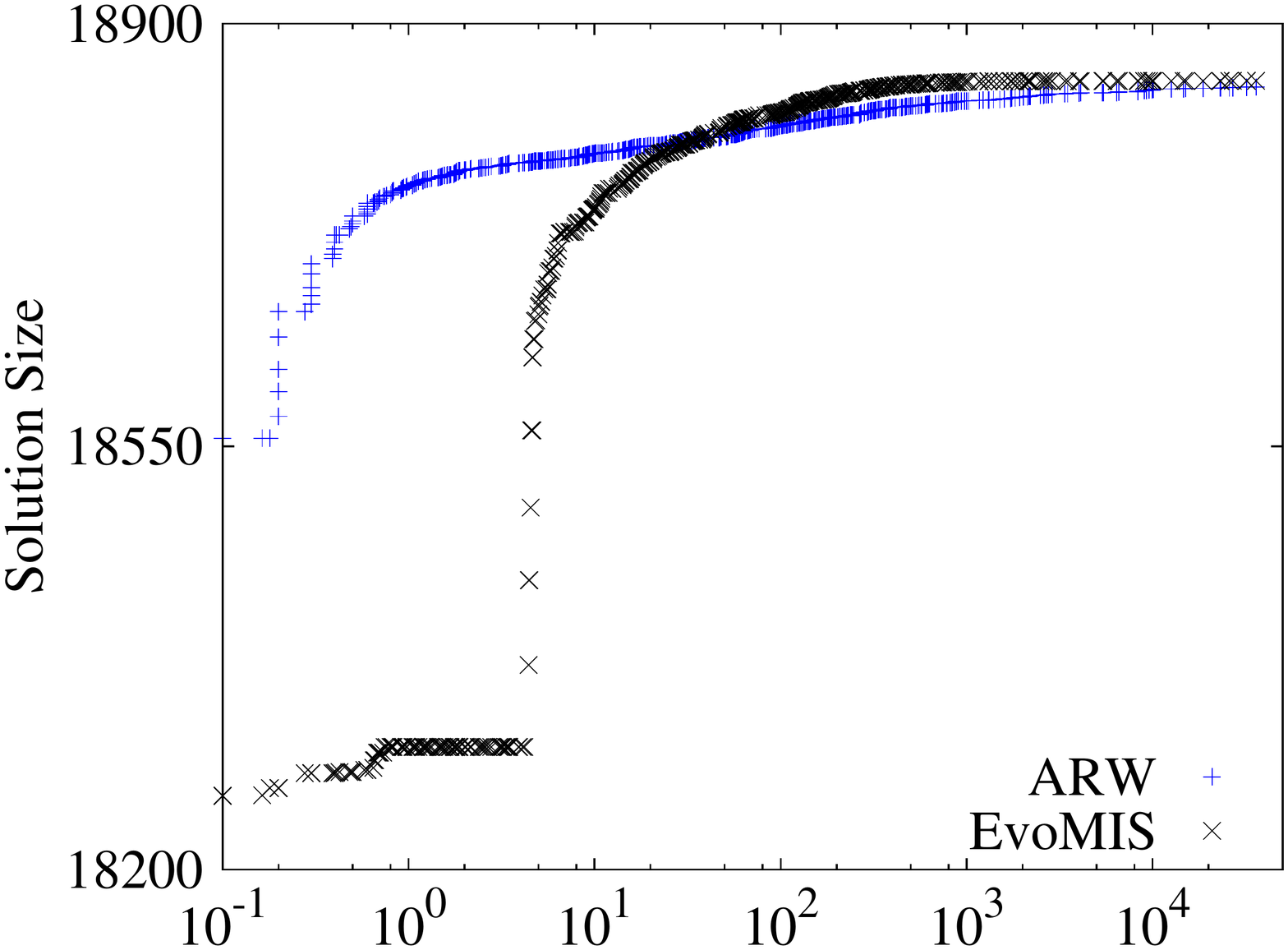}
	\includegraphics[width=8cm]{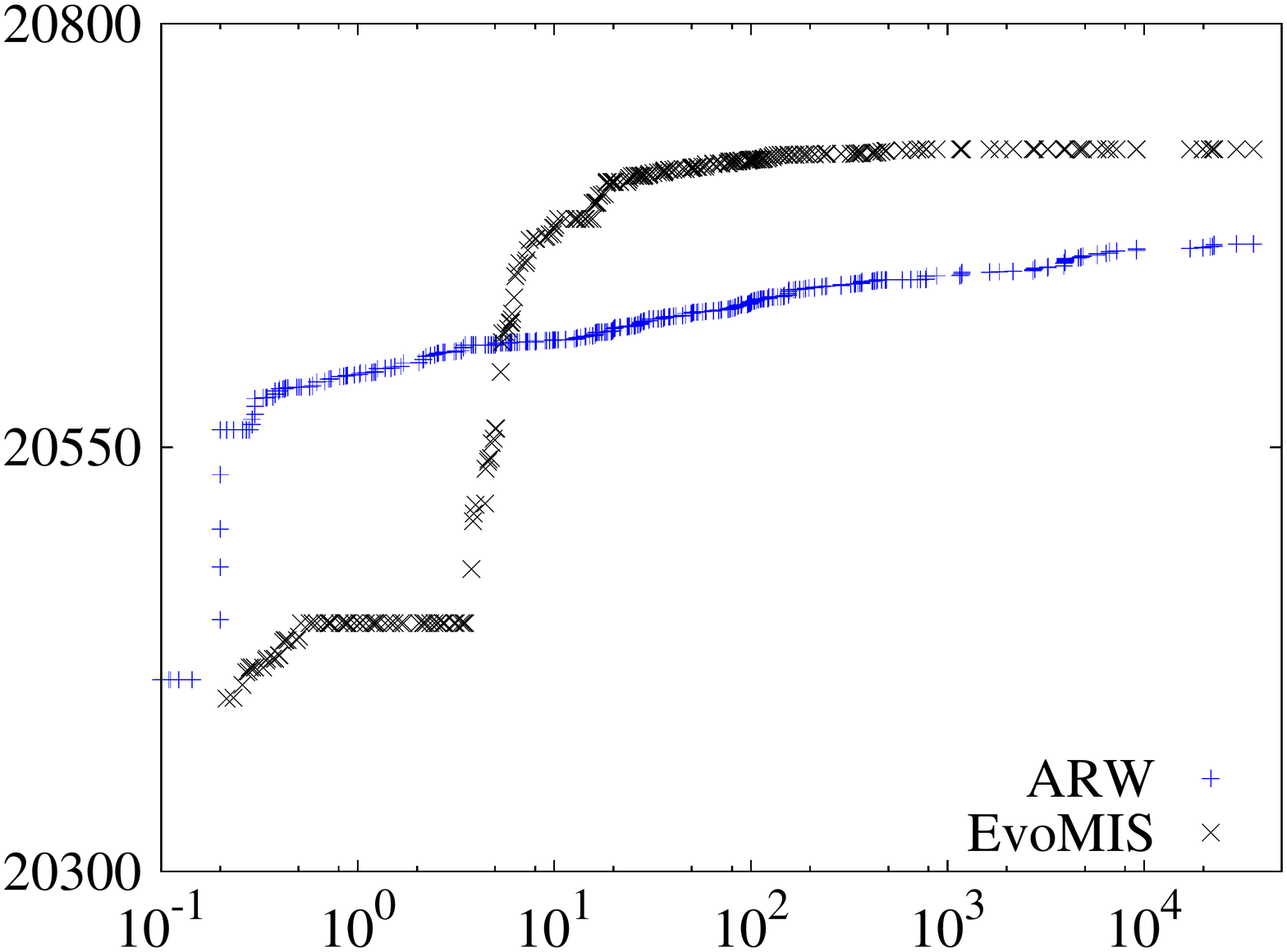} \\
	\includegraphics[width=8cm]{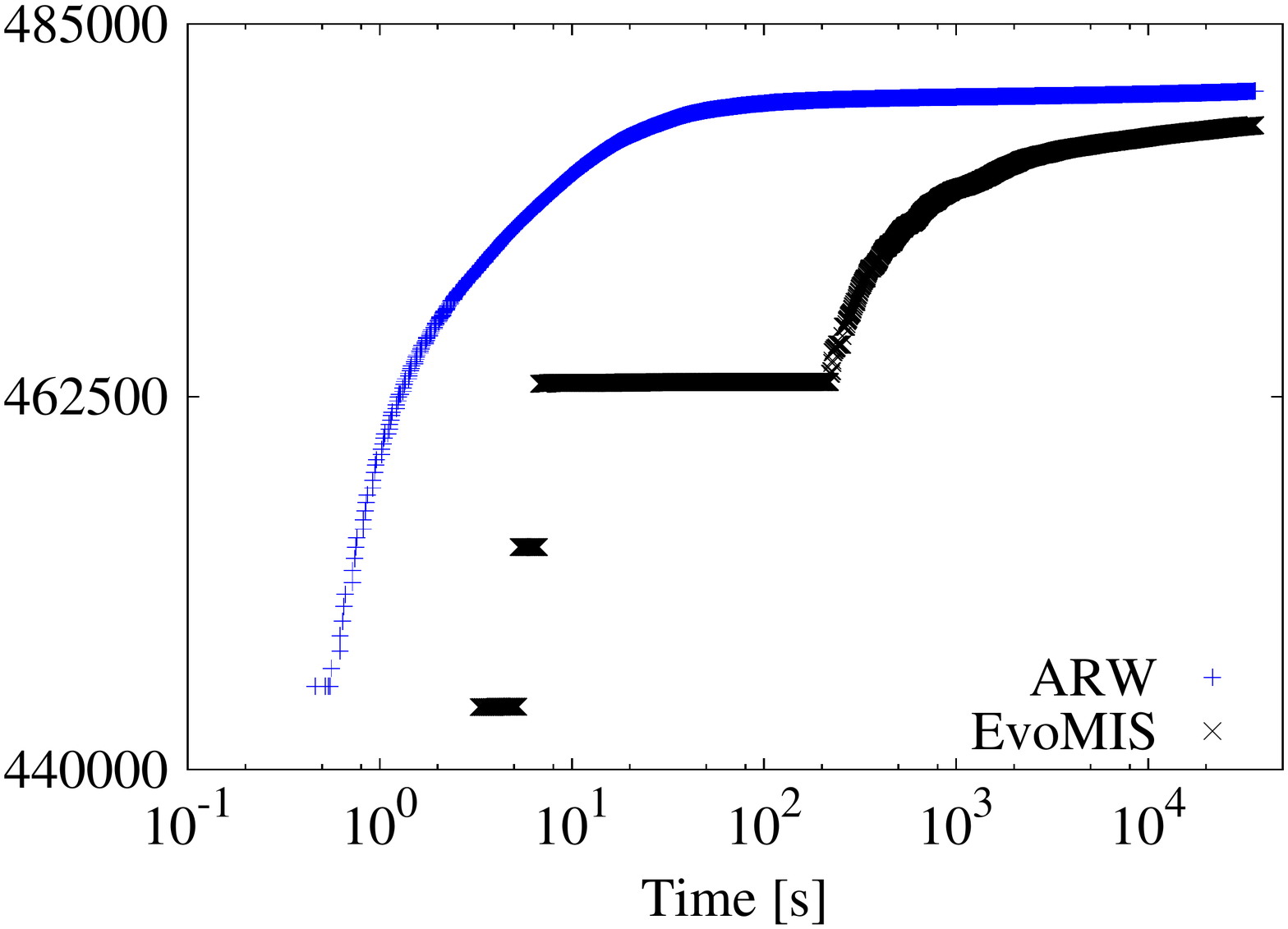}
	\includegraphics[width=8cm]{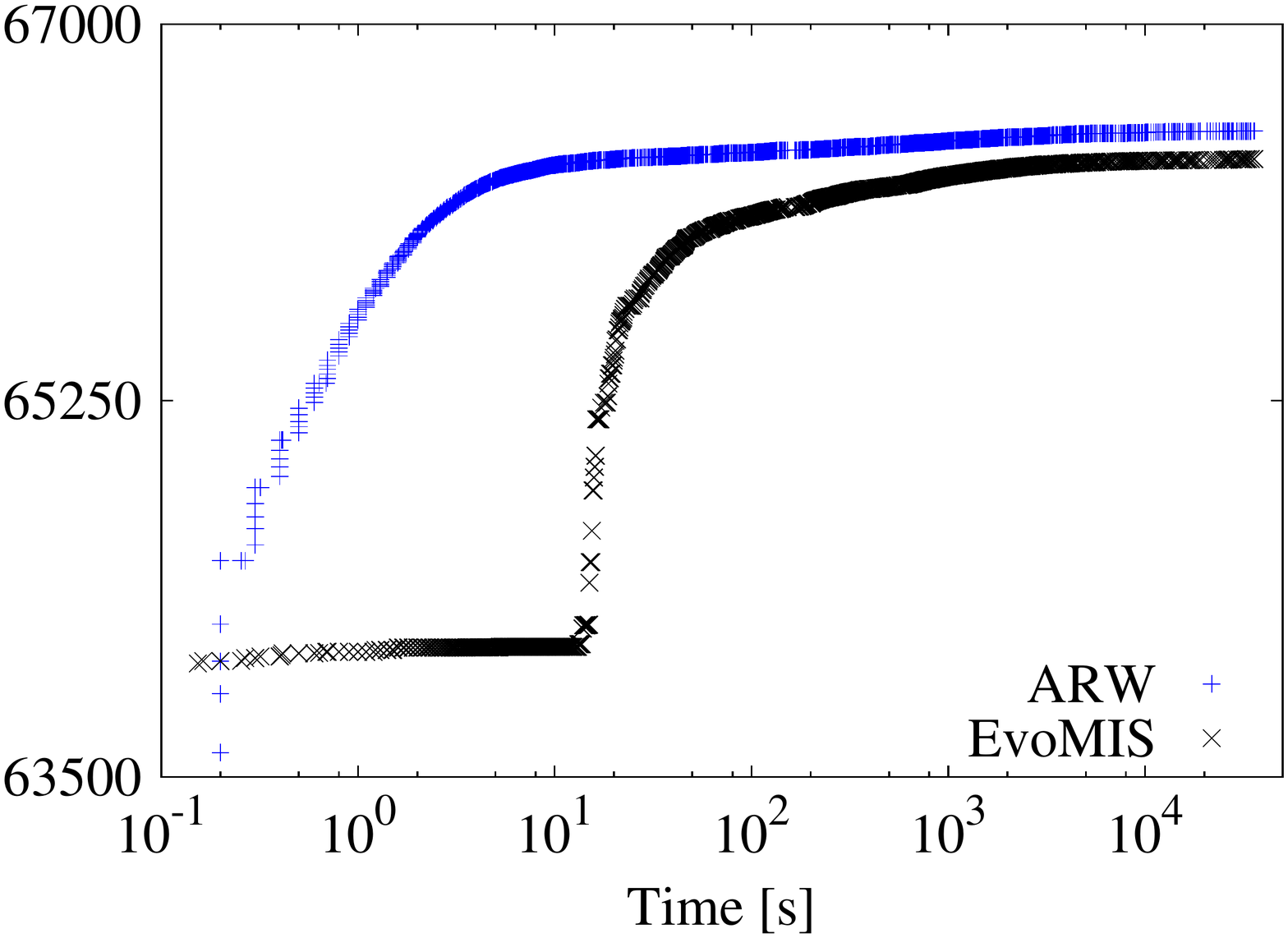} 

        \caption{Solution size evolving over time for four instances from the mesh family: \texttt{feline} and \texttt{gameguy} [top] as well as \texttt{buddha} and \texttt{dragon} [bottom].}
	\label{fig:comp_bunny}
	\label{fig:comp_gameguy}
\end{figure} 

Figure~\ref{fig:comp_bunny} shows how solution quality evolves over time on four example instances from the mesh family for both algorithms.
As one would suspect, our algorithm almost keeps its level of solution quality in the beginning since it has to build the full population before it can start with combine and mutation operations.
Contrarily, the ARW algorithm can directly start with local search and improve its solution.
Hence, the solution quality of the ARW algorithm rises above the solution quality of our algorithm.
As soon as our algorithm finished to compute the population, solution quality starts to improve and eventually the size of the computed independent sets becomes better than the solution quality of the ARW algorithm.

We also implemented the algorithm presented by B\"ack and Khuri~\cite{back1994evolutionary}.
The algorithm uses a two-point crossover as a combine operation, as well as a bit-flip approach for mutation.
Solutions created by the combine and mutation operations can be invalid.
Hence, a penalty approach is used to deal with invalid solution candidates.
In the original paper, the algorithm is only tested on small synthetic or random instances ($\leq200$ nodes). 
We tested the algorithm on the four smallest graphs from the mesh family and gave the algorithm ten hours of time to compute a solution. 
However, the best valid solution created during the course of the algorithm \emph{never} exceeded the size of the best solution after the initial population has been created.
This is due to the fact that the two-point crossover and the mutation operations found valid solutions very rarely so that the average solution quality of the population degrades over time. 
On average, final solution quality of the algorithm has been more than 20\% worse than the final result of our algorithm.
Due to the bad solution quality observed, we did not perform additional experiments with this algorithm.

\paragraph{The Role of Graph Partitioning.}
To estimate the influence of good partitionings in this context, we performed an experiment in which partitions of the graph have been obtained by simple breadth first searches. 
More precisely, we obtain a two-way partition of the graph using a breadth first search starting from a random node.
Every node touched by the breadth first search is added to the first block, and every node not touched by the breadth first search is added to the second block.
The breadth first search is stopped as soon as a specified number of nodes has been touched.
In our experiments, using this approach instead of the approach that uses a graph partitioner to compute a partition yields significantly worse results.
The influence of all the different combine operators that we use here is presented in the thesis~\cite{baLamm}.

\section{Conclusion}
\label{s:conclusion}
We presented a very natural evolutionary framework for the computation of large maximal independent sets. 
Our core innovations are combine operations that are based on graph partitioning and local search algorithms.
More precisely, our combine operations enable us to quickly \emph{exchange whole blocks} of given individuals.
In contrast to previous evolutionary algorithms for the problem, our operators are able to guarantee that the created offspring is valid.  
Experiments indicate that our algorithms outperforms state-of-the-art algorithms on large variety of instances -- some of which are better than every reported in literature.
Important future work includes a coarse-grained parallelization of our approach which can be done by using an island-based approach. 
Moreover, it would be interesting to improve the solution quality of our approach on road networks and to compare our algorithms with exact approaches. 
Additionally, it would be interesting to overcome the slow start of our algorithm due to the initialization of the population.
For example, one could try to adjust the size of the population dynamically.
\section*{Acknowledgements}
We would like to thank Renato Werneck for providing us the source code of the local search algorithms presented in Andrade \etal\cite{AndradeRW12}.
Moreover, we thank the Steinbuch Centre of Computing for giving us access to the IC2 machine. 

\bibliographystyle{plain}
\bibliography{phdthesiscs}

\begin{appendix}
\section{Detailed per Instance Results}
\label{sec:appendixdetailedresults}
\begin{table}[htb]
\centering
\small
\begin{tabular}{lr|rrr|rrr}
\hline
\multicolumn{2}{c}{Graph} & \multicolumn{3}{|c|}{EvoMIS} & \multicolumn{3}{c}{ARW} \\
\hline
Name & $n$ & Avg. & Max. & Min. & Avg. & Max. & Min. \\ 
\hline
\texttt{\detokenize{enron}}    & \numprint{69244}   & \numprint{62811}  & \textbf{\numprint{62811}}  & \numprint{62811}  & \numprint{62811}           & \textbf{\numprint{62811}}  & \numprint{62811}\\
\texttt{\detokenize{gowalla}}  & \numprint{196591}  & \numprint{112369} & \textbf{\numprint{112369}} & \numprint{112369} & \numprint{112369}          & \textbf{\numprint{112369}} & \numprint{112369}\\
\texttt{\detokenize{citation}} & \numprint{268495}  & \numprint{150380} & \textbf{\numprint{150380}} & \numprint{150380} & \numprint{150380} & \textbf{\numprint{150380}} & \numprint{150380}\\
\texttt{\detokenize{cnr-2000}}* & \numprint{325557}  & \numprint{229981} & \textbf{\numprint{229991}} & \numprint{229976} & \numprint{229955}          & \numprint{229966} & \numprint{229940}\\
\texttt{\detokenize{google}}   & \numprint{356648}  & \numprint{174072} & \textbf{\numprint{174072}} & \numprint{174072} & \numprint{174072}          & \textbf{\numprint{174072}} & \numprint{174072}\\
\texttt{\detokenize{coPapers}} & \numprint{434102}  & \numprint{47996}  & \textbf{\numprint{47996}}  & \numprint{47996}  & \numprint{47996}           & \textbf{\numprint{47996}}  & \numprint{47996}\\
\texttt{\detokenize{skitter}}*  & \numprint{554930}  & \numprint{328519} & \numprint{328520}          & \numprint{328519} & \numprint{328609} & \textbf{\numprint{328619}} & \numprint{328599}\\
\texttt{\detokenize{amazon}}   & \numprint{735323}  & \numprint{309774} & \numprint{309778}          & \numprint{309769} & \numprint{309792} & \textbf{\numprint{309793}} & \numprint{309791}\\
\texttt{\detokenize{in-2004}}*  & \numprint{1382908} & \numprint{896581} & \textbf{\numprint{896585}} & \numprint{896580} & \numprint{896477}          & \numprint{896562} & \numprint{896408}\\
\hline
\end{tabular}
\caption{Results for social networks.}
\label{tab:social_network_results}
\end{table}
\begin{table}[htb]
\centering
\small
\begin{tabular}{lr|rrr|rrr}
\hline
\multicolumn{2}{c}{Graph} & \multicolumn{3}{|c|}{EvoMIS} & \multicolumn{3}{c}{ARW} \\
\hline
Name & $n$ & Avg. & Max. & Min. & Avg. & Max. & Min. \\ 
\hline
\texttt{\detokenize{beethoven}} & \numprint{4419}    & \numprint{2004}   & \textbf{\numprint{2004}}  & \numprint{2004}   & \numprint{2004}   & \textbf{\numprint{2004}}    & \numprint{2004}  \\
\texttt{\detokenize{cow}}       & \numprint{5036}    & \numprint{2346}   & \textbf{\numprint{2346}}  & \numprint{2346}   & \numprint{2346}   & \textbf{\numprint{2346}}             & \numprint{2346}  \\
\texttt{\detokenize{venus}}     & \numprint{5672}    & \numprint{2684}   & \textbf{\numprint{2684}}  & \numprint{2684}   & \numprint{2684}   & \textbf{\numprint{2684}}    & \textit{\numprint{2684}}    \\
\texttt{\detokenize{fandisk}}   & \numprint{8634}    & \numprint{4075}   & \textbf{\numprint{4075}}  & \numprint{4075}   & \numprint{4073}   & \numprint{4074}             & \numprint{4072}  \\
\texttt{\detokenize{blob}}      & \numprint{16068}   & \numprint{7249}   & \textbf{\numprint{7250}}  & \numprint{7248}   & \numprint{7249}   & \textbf{\numprint{7250}}    & \numprint{7249}  \\
\texttt{\detokenize{gargoyle}}  & \numprint{20000}   & \numprint{8853}   & \textbf{\numprint{8854}}  & \numprint{8852}   & \numprint{8852}   & \numprint{8853}             & \numprint{8852}  \\
\texttt{\detokenize{face}}      & \numprint{22871}   & \numprint{10218}  & \textbf{\numprint{10218}} & \numprint{10218}  & \numprint{10217}  & \numprint{10217}            & \numprint{10217}  \\
\texttt{\detokenize{feline}}    & \numprint{41262}   & \numprint{18853}  & \textbf{\numprint{18854}} & \numprint{18851}  & \numprint{18847}  & \numprint{18848}            & \numprint{18846}  \\
\texttt{\detokenize{gameguy}}   & \numprint{42623}   & \numprint{20726}  & \textbf{\numprint{20727}} & \numprint{20726}  & \numprint{20670}  & \numprint{20690}            & \numprint{20659}  \\
\texttt{\detokenize{bunny}}*     & \numprint{68790}   & \numprint{32337}  & \textbf{\numprint{32343}} & \numprint{32330}  & \numprint{32293}  & \numprint{32300}            & \numprint{32287}  \\
\texttt{\detokenize{dragon}}    & \numprint{150000}  & \numprint{66373}  & \numprint{66383}          & \numprint{66365}  & \numprint{66503}  & \textbf{\numprint{66505}}            & \numprint{66500}  \\
\texttt{\detokenize{turtle}}    & \numprint{267534}  & \numprint{122378} & \numprint{122391}         & \numprint{122370} & \numprint{122506} & \textbf{\numprint{122584}}           & \numprint{122444} \\
\texttt{\detokenize{dragonsub}} & \numprint{600000}  & \numprint{281403} & \numprint{281436}         & \numprint{281384} & \numprint{282006} & \textbf{\numprint{282066}}           & \numprint{281954} \\
\texttt{\detokenize{ecat}}      & \numprint{684496}  & \numprint{322285} & \numprint{322357}         & \numprint{322222} & \numprint{322362} & \textbf{\numprint{322529}}           & \numprint{322269} \\
\texttt{\detokenize{buddha}}    & \numprint{1087716} & \numprint{478879} & \numprint{478936}         & \numprint{478795} & \numprint{480942} & \textbf{\numprint{480969}} & \numprint{480921} \\
\hline
\end{tabular}
\caption{Results for mesh type graphs.}
\label{tab:mesh_results}
\end{table}

\begin{table}[htb]
\centering
\small
\begin{tabular}{lr|rrr|rrr}
\hline
\multicolumn{2}{c}{Graph} & \multicolumn{3}{|c|}{EvoMIS} & \multicolumn{3}{c}{ARW} \\
\hline
Name & $n$ & Avg. & Max. & Min. & Avg. & Max. & Min. \\ 
\hline
\texttt{\detokenize{crack}}    & \numprint{10240}  & \numprint{4603}    & \textbf{\numprint{4603}}  & \numprint{4603}  & \numprint{4603}  & \textbf{\numprint{4603}}  & \numprint{4603} \\
\texttt{\detokenize{vibrobox}} & \numprint{12328}  & \numprint{1852}    & \textbf{\numprint{1852}}  & \numprint{1852}  & \numprint{1850}  & \numprint{1851}  & \numprint{1849} \\
\texttt{\detokenize{4elt}}     & \numprint{15606}  & \numprint{4944}    & \textbf{\numprint{4944}}  & \numprint{4944}  & \numprint{4942}  & \textbf{\numprint{4944}}  & \numprint{4940} \\
\texttt{\detokenize{cs4}}      & \numprint{22499}  & \numprint{9172}    & \textbf{\numprint{9177}}  & \numprint{9170}  & \numprint{9173}  & \numprint{9174}  & \numprint{9172} \\
\texttt{\detokenize{bcsstk30}} & \numprint{28924}  & \numprint{1783}    & \textbf{\numprint{1783}}  & \numprint{1783}  & \numprint{1783}  & \textbf{\numprint{1783}}  & \numprint{1783} \\
\texttt{\detokenize{bcsstk31}} & \numprint{35588}  & \numprint{3488}    & \textbf{\numprint{3488}}  & \numprint{3488}  & \numprint{3487}  & \numprint{3487}  & \numprint{3487} \\
\texttt{\detokenize{fe_pwt}}   & \numprint{36519}  & \numprint{9309}  & \textbf{\numprint{9310}}  & \numprint{9309}  & \numprint{9310}  & \textbf{\numprint{9310}}  & \numprint{9308} \\
\texttt{\detokenize{brack2}}   & \numprint{62631}  & \numprint{21417}   & \textbf{\numprint{21417}} & \numprint{21417} & \numprint{21416} & \numprint{21416} & \numprint{21415} \\
\texttt{\detokenize{fe_tooth}} & \numprint{78136}  & \numprint{27793}   & \textbf{\numprint{27793}} & \numprint{27793} & \numprint{27792} & \numprint{27792} & \numprint{27791} \\
\texttt{\detokenize{fe_rotor}} & \numprint{99617}  & \numprint{22022} & \numprint{22026}          & \numprint{22019} & \numprint{21974} &\textbf{\numprint{22030}} & \numprint{21902} \\
\texttt{\detokenize{598a}}     & \numprint{110971} & \numprint{21826}   & \numprint{21829}          & \numprint{21824} & \numprint{21891} & \textbf{\numprint{21894}} & \numprint{21888} \\
\texttt{\detokenize{wave}}     & \numprint{156317} & \numprint{37057} & \textbf{\numprint{37063}} & \numprint{37046} & \numprint{37023} & \numprint{37040} & \numprint{36999} \\
\texttt{\detokenize{fe_ocean}} & \numprint{143437} & \numprint{71390}   & \numprint{71576}          & \numprint{71233} & \numprint{71492} & \textbf{\numprint{71655}} & \numprint{71291} \\
\texttt{\detokenize{auto}}     & \numprint{448695} & \numprint{83935} & \numprint{83969}          & \numprint{83907} & \numprint{84462} & \textbf{\numprint{84478}} & \numprint{84453} \\
\hline
\end{tabular}
\caption{Results for Walshaw benchmark graphs.}
\label{tab:walshaw_results}
\end{table}
\begin{table}[htb]
\centering
\small
\begin{tabular}{lr|rrr|rrr}
\hline
\multicolumn{2}{c}{Graph} & \multicolumn{3}{|c|}{EvoMIS} & \multicolumn{3}{c}{ARW} \\
\hline
Name & $n$ & Avg. & Max. & Min. & Avg. & Max. & Min. \\ 
\hline
\texttt{\detokenize{ny}}  & \numprint{264346}  & \numprint{131384} & \numprint{131395} & \numprint{131377} &\numprint{131481} & \textbf{\numprint{131485}} & \numprint{131476} \\
\texttt{\detokenize{bay}} & \numprint{321270}  & \numprint{166329} & \numprint{166345} & \numprint{166318} &\numprint{166368} & \textbf{\numprint{166375}} & \numprint{166364}  \\
\texttt{\detokenize{col}} & \numprint{435666}  & \numprint{225714} & \numprint{225721} & \numprint{225706} &\numprint{225764} & \textbf{\numprint{225768}} & \numprint{225759}  \\
\texttt{\detokenize{fla}} & \numprint{1070376} & \numprint{549093} & \numprint{549106} & \numprint{549072} &\numprint{549581} & \textbf{\numprint{549587}} & \numprint{549574}  \\
\hline
\end{tabular}
\caption{Results for road networks.}
\label{tab:street_network_results}
\end{table}
\begin{table}[htb]
\centering
\small
\begin{tabular}{lr|rrr|rrr}
\hline
\multicolumn{2}{c}{Graph} & \multicolumn{3}{|c|}{EvoMIS} & \multicolumn{3}{c}{ARW} \\
\hline
Name & $n$ & Avg. & Max. & Min. & Avg. & Max. & Min. \\ 
\hline
\texttt{\detokenize{Oregon-1}} & \numprint{11174} & \numprint{9512}  & \textbf{\numprint{9512}}  & \numprint{9512}  & \numprint{9512}  & \textbf{\numprint{9512}}  & \numprint{9512}       \\
\texttt{\detokenize{ca-HepPh}} & \numprint{12006} & \numprint{4994}  & \textbf{\numprint{4994}}  & \numprint{4994}  & \numprint{4994}  & \textbf{\numprint{4994}}  & \numprint{4994}       \\
\texttt{\detokenize{skirt}}    & \numprint{12595} & \numprint{2383}     & \textbf{\numprint{2383}}              & \numprint{2383}     & \numprint{2383}  & \textbf{\numprint{2383}}           & \numprint{2383}       \\
\texttt{\detokenize{cbuckle}}  & \numprint{13681} & \numprint{1097}  & \textbf{\numprint{1097}}  & \numprint{1097}  & \numprint{1097}  & \textbf{\numprint{1097}}  & \numprint{1097}       \\
\texttt{\detokenize{cyl6}}     & \numprint{13681} & \numprint{600}   & \textbf{\numprint{600}}   & \numprint{600}   & \numprint{600}   & \textbf{\numprint{600}}   & \numprint{600}        \\
\texttt{\detokenize{case9}}    & \numprint{14453} & \numprint{7224}  & \textbf{\numprint{7224}}  & \numprint{7224}  & \numprint{7224}  & \textbf{\numprint{7224}}  & \numprint{7224}       \\
\texttt{\detokenize{rajat07}}  & \numprint{14842} & \numprint{4971}  & \textbf{\numprint{4971}}  & \numprint{4971}  & \numprint{4971}  & \textbf{\numprint{4971}}  & \numprint{4971}       \\
\texttt{\detokenize{Dubcova1}} & \numprint{16129} & \numprint{4096}  & \textbf{\numprint{4096}}  & \numprint{4096}  & \numprint{4096}  & \textbf{\numprint{4096}}  & \numprint{4096}       \\
\texttt{\detokenize{olafu}}    & \numprint{16146} & \numprint{735}   & \textbf{\numprint{735}}   & \numprint{735}   & \numprint{735}   & \textbf{\numprint{735}}   & \numprint{735}        \\
\texttt{\detokenize{bodyy6}}   & \numprint{19366} & \numprint{6232}  & \textbf{\numprint{6233}}  & \numprint{6230}  & \numprint{6226}  & \numprint{6228}           & \numprint{6224}        \\
\texttt{\detokenize{raefsky4}} & \numprint{19779} & \numprint{1055}  & \textbf{\numprint{1055}}  & \numprint{1055}  & \numprint{1053}  & \numprint{1053}           & \numprint{1053}       \\
\texttt{\detokenize{smt}}      & \numprint{25710} & \numprint{782}   & \textbf{\numprint{782}}   & \numprint{782}   & \numprint{780}   & \numprint{780}            & \numprint{780}        \\
\texttt{\detokenize{pdb1HYS}}  & \numprint{36417} & \numprint{1078}  & \textbf{\numprint{1078}}  & \numprint{1078}  & \numprint{1070}  & \numprint{1071}           & \numprint{1070}       \\
\texttt{\detokenize{c-57}}     & \numprint{37833} & \numprint{19997} & \textbf{\numprint{19997}} & \numprint{19997} & \numprint{19997} & \textbf{\numprint{19997}} & \numprint{19997}      \\
\texttt{\detokenize{copter2}}          & \numprint{55476} & \numprint{15192} & \textbf{\numprint{15195}} & \numprint{15191} & \numprint{15186} & \numprint{15194} & \numprint{15179}\\
\texttt{\detokenize{TSOPF_FS_b300_c2}} & \numprint{56813} & \numprint{28338}   & \textbf{\numprint{28338}} & \numprint{28338} & \numprint{28338}   & \textbf{\numprint{28338}} & \numprint{28338}\\
\texttt{\detokenize{c-67}}             & \numprint{57975} & \numprint{31257}   & \textbf{\numprint{31257}} & \numprint{31257} & \numprint{31257}   & \textbf{\numprint{31257}} & \numprint{31257}\\
\texttt{\detokenize{dixmaanl}}         & \numprint{60000} & \numprint{20000}   & \textbf{\numprint{20000}} & \numprint{20000} & \numprint{20000}   & \textbf{\numprint{20000}} & \numprint{20000}\\
\texttt{\detokenize{blockqp1}}         & \numprint{60012} & \numprint{20011}   & \textbf{\numprint{20011}} & \numprint{20011} & \numprint{20011}   &\textbf{\numprint{20011}} & \numprint{20011}\\
\texttt{\detokenize{Ga3As3H12}}        & \numprint{61349} & \numprint{8118}  & \textbf{\numprint{8151}}  & \numprint{8097}  & \numprint{8061}  & \numprint{8124}  & \numprint{7842}\\
\texttt{\detokenize{GaAsH6}}           & \numprint{61349} & \numprint{8562}  & \numprint{8572}           & \numprint{8547}  & \numprint{8519}  & \textbf{\numprint{8575}}  & \numprint{8351}\\
\texttt{\detokenize{cant}}             & \numprint{62208} & \numprint{6260}    & \textbf{\numprint{6260}}  & \numprint{6260}  & \numprint{6255}  & \numprint{6255}  & \numprint{6254}\\
\texttt{\detokenize{ncvxqp5}}          & \numprint{62500} & \numprint{24526} & \numprint{24537}          & \numprint{24510} & \numprint{24580} & \textbf{\numprint{24608}} & \numprint{24520}\\
\texttt{\detokenize{crankseg_2}}       & \numprint{63838} & \numprint{1735}    & \textbf{\numprint{1735}}  & \numprint{1735}  & \numprint{1735}    & \textbf{\numprint{1735}}  & \numprint{1735}\\
\texttt{\detokenize{c-68}}             & \numprint{64810} & \numprint{36546}   & \textbf{\numprint{36546}} & \numprint{36546} & \numprint{36546}   & \textbf{\numprint{36546}} & \numprint{36546}\\
\hline
\end{tabular}
\caption{Results for graphs from Florida Sparse Matrix collection.}
\label{tab:mesh_results}
\end{table}

\end{appendix}
\end{document}

%% file: makros.tex
\usepackage{amsfonts}

\newcommand{\set}[1]{\left\{ #1\right\}}
\newcommand{\gilt}{:}
\newcommand{\sodass}{\,:\,}
\newcommand{\setGilt}[2]{\left\{ #1\sodass #2\right\}}

\newcommand{\realrange}[2]{\left[#1, #2\right]}

\newcommand{\unitrange}[2]{\realrange{0}{1}}

\newcommand{\llabel}[1]{\label{\labelprefix:#1}}
\newcommand{\labelprefix}{} 

\newcommand{\discussionsize}{\small}

\newcommand{\frage}[1]{}

\newenvironment{code}{\noindent
\begin{tabbing}%
\hspace{2em}\=\hspace{2em}\=\hspace{2em}\=\hspace{2em}\=\hspace{2em}\=%
\hspace{2em}\=\hspace{2em}\=\hspace{2em}\=\hspace{2em}\=\hspace{2em}\=%
\kill}{\end{tabbing}}

\newcommand{\labelcommand}{}
\newcommand{\captiontext}{}
\newsavebox{\codeparam}
\newcounter{lineNumber}
\newenvironment{disscodepos}[3]{%
\renewcommand{\labelcommand}{#2}%
\renewcommand{\captiontext}{#3}%
\sbox{\codeparam}{\parbox{\textwidth}{#3}}%
\begin{figure}[#1]\begin{center}\begin{code}\setcounter{lineNumber}{1}}{%
\end{code}\end{center}\caption{\llabel{\labelcommand}\captiontext}\end{figure}}

{\end{disscodepos}}

\newcommand{\Is}       {:=}

\newdimen\endofsize\endofsize=0.5em
\def\endofbeweis{~\quad\hglue\hsize minus\hsize
                 \hbox{\vrule height \endofsize width
\endofsize}\par}
